\documentclass{emulateapj}
\usepackage{mathptmx}
\usepackage{graphicx}

\begin{document}
\title{Type Ia Supernova science 2010 -- 2020}

\author{D.~A.~Howell\altaffilmark{1,2},
A.~Conley\altaffilmark{3},
M.~Della Valle\altaffilmark{4},
P.~E.~Nugent\altaffilmark{5},
S.~Perlmutter\altaffilmark{5,6},
G.~H.~Marion\altaffilmark{7},
K.~Krisciunas\altaffilmark{8},
C.~Badenes\altaffilmark{9},
P.~Mazzali\altaffilmark{10,11},
G.~Aldering\altaffilmark{5},
P.~Antilogus\altaffilmark{12},
E.~Baron\altaffilmark{13},
A.~Becker\altaffilmark{14},
C.~Baltay\altaffilmark{15},
S.~Benetti\altaffilmark{16},
S.~Blondin\altaffilmark{17},
D.~Branch\altaffilmark{13},
E.~F.~Brown\altaffilmark{18},
S.~Deustua\altaffilmark{19},
A.~Ealet\altaffilmark{20},
R.~S.~Ellis\altaffilmark{21,22},
D.~Fouchez\altaffilmark{12},
W.~Freedman\altaffilmark{23},
A.~Gal-Yam\altaffilmark{24},
S.~Jha\altaffilmark{25},
D.~Kasen\altaffilmark{26},
R.~Kessler\altaffilmark{27},
A.~G.~Kim\altaffilmark{5},
D.~C.~Leonard\altaffilmark{28},
W.~Li\altaffilmark{6},
M.~Livio\altaffilmark{19},
D.~Maoz\altaffilmark{29},
F.~Mannucci\altaffilmark{30},
T.~Matheson\altaffilmark{31},
J.~D.~Neill\altaffilmark{22},
K.~Nomoto\altaffilmark{32},
N.~Panagia\altaffilmark{19,33,34},
K.~Perrett\altaffilmark{35},
M.~Phillips\altaffilmark{36,37},
D.~Poznanski\altaffilmark{6},
R.~Quimby\altaffilmark{22},
A.~Rest\altaffilmark{38},
A.~Riess\altaffilmark{19,39},
M.~Sako\altaffilmark{40},
A.~M.~Soderberg\altaffilmark{41},
L.~Strolger\altaffilmark{42},
R.~Thomas\altaffilmark{5},
M.~Turatto\altaffilmark{33},
S.~van Dyk\altaffilmark{43},
W.~M.~Wood-Vasey\altaffilmark{44}
}

\altaffiltext{1}{Las Cumbres Observatory Global Telescope Network}
\altaffiltext{2}{University of California, Santa Barbara}
\altaffiltext{3}{University of Colorado}         
\altaffiltext{4}{INAF - Capodimonte Observatory; International Center for Relativistic Astrophysics}
\altaffiltext{5}{Lawrence Berkeley National Laboratory}
\altaffiltext{6}{University of California, Berkeley}
\altaffiltext{7}{University of Texas}
\altaffiltext{8}{Texas A \& M University}
\altaffiltext{9}{Princeton University}
\altaffiltext{10}{MPA}
\altaffiltext{11}{INAF Osservatorio Astronomico di Trieste}
\altaffiltext{12}{CNRS-IN2P3 and Universit\'e Paris VI \& VII}
\altaffiltext{13}{University of Oklahoma}
\altaffiltext{14}{University of Washington}
\altaffiltext{15}{Yale University}
\altaffiltext{16}{Istituto Nazionale di Astrofisica - Osservatorio Astronomico di Padova}    
\altaffiltext{17}{ESO}                        
\altaffiltext{18}{Michigan State University}  
\altaffiltext{19}{Space Telescope Science Institute}                      
\altaffiltext{20}{Centre de Physique des Particules de Marseille}
\altaffiltext{21}{Oxford University}
\altaffiltext{22}{Caltech}                                              
\altaffiltext{23}{Carnegie Observatories}                             
\altaffiltext{24}{Weizmann Institute}                   
\altaffiltext{25}{Rutgers University }                  
\altaffiltext{26}{University of California, Santa Cruz} 
\altaffiltext{27}{KICP}                                 
\altaffiltext{28}{San Diego State University}           
\altaffiltext{29}{Tel-Aviv University}                   
\altaffiltext{30}{INAF - Osservatorio di Arcetri}        
\altaffiltext{31}{NOAO}                                  
\altaffiltext{32}{University of Tokyo}                   
\altaffiltext{33}{INAF-Catania Astrophysical Observatory}
\altaffiltext{34}{Supernova Ltd.} 
\altaffiltext{35}{University of Toronto}    
\altaffiltext{36}{Carnegie Institution of Washington}
\altaffiltext{37}{Las Campanas Observatory}              
\altaffiltext{38}{Harvard University}                    
\altaffiltext{39}{Johns Hopkins University}       
\altaffiltext{40}{University of Pennsylvania}            
\altaffiltext{41}{Harvard/CfA}                           
\altaffiltext{42}{Western Kentucky University}           
\altaffiltext{43}{SSC/Caltech}    
\altaffiltext{44}{University of Pittsburgh}   

\begin{abstract}
  In the next decade Type Ia supernovae (SNe Ia) will be used to test
  theories predicting changes in the Dark Energy equation of state
  with time.  Ultimately this requires a dedicated space mission like
  JDEM.  SNe Ia are mature cosmological probes --- their limitations
  are well characterized, and a path to improvement is clear.
  Dominant systematic errors include photometric calibration,
  selection effects, reddening, and population-dependent differences.
  Building on past lessons, well-controlled new surveys are poised to
  make strides in these areas: the Palomar Transient Factory,
  Skymapper, La Silla QUEST, Pan-STARRS, the Dark Energy Survey, LSST,
  and JDEM.  They will obviate historical calibrations and selection
  biases, and allow comparisons via large subsamples.  Some
  systematics follow from our ignorance of SN Ia progenitors, which
  there is hope of determining with SN Ia rate studies from $0<z<4$.

  Aside from cosmology, SNe Ia regulate galactic and cluster chemical
  evolution, inform stellar evolution, and are laboratories for
  extreme physics.  Essential probes of SNe Ia in these contexts
  include spectroscopy from the UV to the IR, X-ray cluster and SN
  remnant observations, spectropolarimetry, and advanced theoretical
  studies. While there are an abundance of discovery facilities
  planned, there is a deficit of follow-up resources.  Living in the
  systematics era demands deep understanding rather than larger
  statistics.  NOAO ReSTAR initiative to build 2-4m telescopes would
  provide necessary follow-up capability.  Finally, to fully exploit
  LSST, well-matched wide-field spectroscopic capabilities are
  desirable.
\end{abstract}

\section{Introduction}
\begin{figure}
\plotone{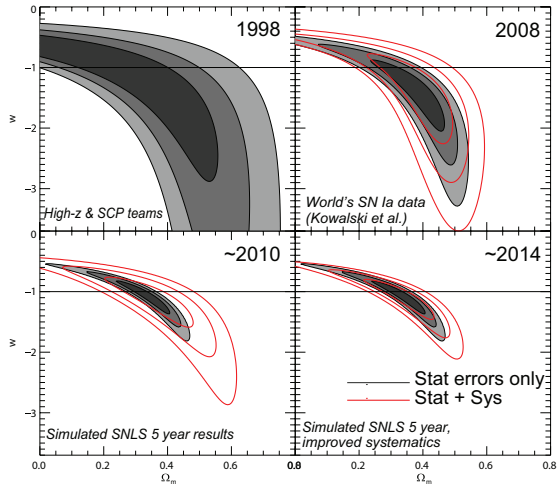} 
\caption{The top two panels show the remarkable improvements in $w$
  made using SNe Ia over the past decade (assuming a flat universe),
  the bottom two show the importance of improving systematics in the
  early years of the next decade.  {\it Upper left:} $\Omega_M - w$
  statistical-only constraints circa 1998 \citep[see
  ][]{1998ApJ...493L..53G}.  {\it Upper right:} By 2008, Kowalski et
  al., combining many data sets, showed that systematic errors are
  significant. {\it Bottom left:} Expected constraints for the year 5
  results of SNLS, assuming additional low-z SNe, and double the
  number of $z > 1$ SNe from HST, and assuming there is no improvement in
  systematic errors from the 3rd year result.  {\it Bottom right:}
  Assumes the low-z data are on the SDSS photometric system, and a
  factor of two improvement in measurements of fundamental flux
  standards.  Judged by the DETF figure of merit (the area of the
  inner 68.3\% contour), the improvement from the 1998 results is a
  factor of 3, 5, and 10, including systematics }
\label{wplot}
\end{figure}

A decade ago, Type Ia supernovae (SNe Ia) were used as standardized
candles to reveal the presence of a previously unknown energy
component of the universe which dominates its evolution
\citep{1998AJ....116.1009R, 1999ApJ...517..565P}.  We now know the
Dark Energy, averaged over cosmic time, behaves similarly to a
cosmological constant, $\langle w=P/\rho c^2 \rangle \simeq -1 \pm
6\%$ (stat, $1\sigma$), with systematic errors of the same order or
larger than statistical errors
\citep[][Fig.~\ref{wplot}]{2008ApJ...686..749K,2009arXiv0901.4804H}.
The challenge for the next decade is to measure the variation of $w$
with redshift.  Excellent progress has been made in identifying errors
that do not scale with $\sqrt{N}$ statistics, i.e. systematic errors
(Table 1), and we now envision
experiments that will address decades-old uncertainties enabling
breakthroughs in the use of SNe Ia as standard candles.  In the next
ten years there is a high probability we will be able to answer the
questions: ``What are the progenitors of SNe Ia?  Why does their
brightness change with lightcurve shape, color, stellar population
age, or metallicity?  Do they evolve with redshift?  What fraction are
aspherical, and why?  What role do SNe Ia play in galactic chemical
enrichment?,'' and ``How does extragalactic dust compare to Milky Way
dust?''

The path to answering most of these questions and improving SNe as
standard candles is the same: construct large subsamples of SNe Ia
split by various properties, and study correlations between them.
This is a paradigm shift that will be enabled by huge new SN discovery
projects.  First we discuss the planned new projects (\S 2),
then we discuss the above scientific questions
\S 3-6.  In \S 7 we discuss the systematic
errors temporarily affecting SN cosmology and prospects for their
eradication, and we conclude with recommendations in \S 8.

\section{Outlook\label{outlook}}
All current SN studies are sample-size limited.  But in the next 10
years, that will no longer be the case.  With thousands of SNe Ia
discovered per year (we will need a new naming convention), we are
leaving the serendipity-driven era, where we learn what nature wants
to tell us, and entering the hypothesis-driven era, where
large-N subsamples can be constructed to test ideas.  We can compare
SNe in ellipticals to those in spirals, split them by color, redshift
bin, ejecta velocity, or host metallicity.  We will be able to
correlate IR or polarimetric properties against spectral features,
optical properties, or host galaxy features.  
We can create data
``cubes'' in dozens of dimensions.  It is hard to imagine where the
most exciting discoveries will come from.

The most obvious progress will be at the high redshift frontier.
At $z>1$, less than two dozen SNe Ia are
known \citep{2007ApJ...659...98R}.  A refurbished HST should use ACS
to continue to build the $1.0 < z < 1.5$ sample necessary to study the
time evolution of $w$.  However, to build a truly large, uniform
sample will require JDEM/EUCLID.  HST WFC3 will provide the first
glimpse of SNe Ia at $1.5 < z < 3$ in the matter-dominated era
\citep{2006ApJ...648..884R}.  The next generation of large telescopes,
JWST, TMT, E-ELT and GMT will allow spectroscopy of $z>3$ SNe, and
extend studies to even higher redshifts.  These studies may at 
last elucidate the progenitors of SNe Ia, and
enable studies of SN Ia evolution over vast stretches of cosmic time.

Currently, intermediate redshift ranges are the best studied --
ESSENCE \citep{2008ApJ...689..377W} and the Supernova Legacy Survey
\citep[SNLS;][]{2006A&A...447...31A} have obtained (but only published
a fraction of) about 600 well measured, spectroscopically confirmed
SNe Ia at $0.1<z<1.0$.  The SDSS \citep{2008AJ....136.2306H} has $\sim
500$ spectroscopically confirmed supernovae at $0.05<z<0.3$.  Over the
next few years the Pan-STARRS Medium Deep Survey (MDS) and the Dark
Energy Survey will discover thousands of supernovae at $0.1<z<1.0$,
though they will only be able to spectroscopically confirm a fraction
of them.  Ultimately LSST will
produce tens of thousands of well measured SNe Ia per year in this
redshift range, but will be limited by the number of follow-up
facilities available.  Many SN studies are done most efficiently at
intermediate redshifts where large numbers of SNe can be studied over
a few square degrees using a rolling search.

At low redshift SNe Ia have, until this point, been studied in more of a
piecemeal fashion, necessitated by the lack of a multiplex advantage.
Thus they suffer from, and cause, some of the largest
systematic errors affecting SNe Ia (\S\ref{sys}).  However, in this
regime there may be the greatest room for optimism in
the coming decade.  Within a few years, programs already underway,
KAIT \citep{2001ApJ...546..734L}, the Carnegie SN Program
\citep[CSP; ][]{2006PASP..118....2H}, the CfA program
\citet{2009arXiv0901.4804H}, and the Nearby SN Factory
\citep{2006ApJ...650..510A} will produce a total of $\gtrsim 300$
cosmologically useful $z<0.1$ SNe Ia.  By 2010 a new wave of dedicated
multi-square-degree detectors on small telescopes will be available,
each of which will discover hundreds of SNe per year: Skymapper
\citep[5.7 sq. deg.  FOV;][]{2007PASA...24....1K}, the Palomar
Transient Factory \citep[7.5 sq. deg. FOV;][]{2008SPIE.7014E.163R},
the La Silla SN search (the QUEST camera on the La Silla Schmidt
telescope), and the Pan-STARRS 3$\pi$ search (7 sq. deg. FOV).  

Low redshifts are where the fundamental work for understanding SNe Ia
is done --- time series spectroscopy, space-based UV follow-up,
ground-based IR photometry and spectra, and spectropolarimetry.
Because many of the systematic errors limiting SN Ia cosmology are a
result of astrophysical ignorance, strides in this regime have an
impact across SN cosmology at all redshifts, as seen in the bottom
panels of Fig.~\ref{wplot}.

\section{The progenitor question\label{progen}}
There is consensus that SNe Ia are the result of the explosion
of a carbon-oxygen white dwarf that grows to near the Chandrasekhar
limit in a binary system \citep{1960ApJ...132..565H}.  But is debate
over whether the companion is an evolved or main sequence star
\citep[single degenerate system; ][]{1973ApJ...186.1007W}, or whether
it is another white dwarf, i.e. a double
degenerate system \citep{1984ApJS...54..335I,1984ApJ...277..355W}.
The two scenarios produce different delay times from the birth of the
binary system to explosion, so there is hope of deducing the
progenitors of SNe Ia by studying their delay time distribution (DTD).

Empirically, the delay time distribution can be determined from the
lag between the cosmic star formation rate and the SN Ia birthrate.
However, without exquisite data, the shape can be ambiguous, with
authors advocating a single Gaussian delay time
\citep{2004MNRAS.347..942G,2004ApJ...613..200S}, bimodality,
\citep{2006MNRAS.370..773M, 2005ApJ...629L..85S, 2006ApJ...648..868S}
or a continuous, declining DTD from young to old stellar ages
\citep{2008ApJ...683L..25P,2008PASJ...60.1327T}.  These DTDs can then
be compared to theoretical models to determine the progenitor
\citep{2005ApJ...629..915B, 2005A&A...441.1055G, 2008ApJ...683L..25P,
  2008ApJ...679.1390H, 2008ApJ...683L.127H}

\begin{figure}
\plotone{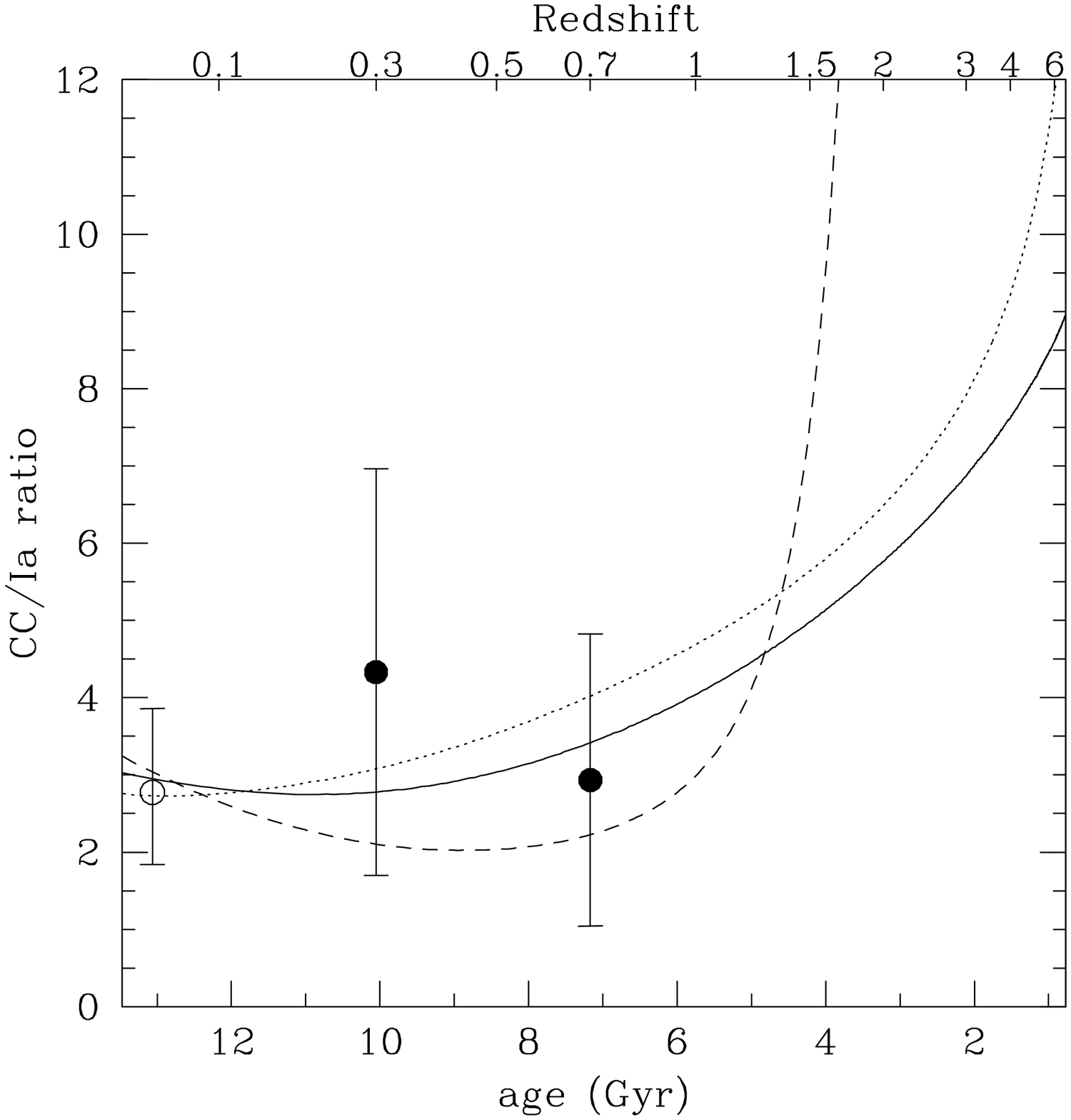} 
\caption{ Ratio of the rates of the CC to Ia SNe as a function of
  redshift.  The white and black dots are observed values
  \citep{2004ApJ...613..189D,2005A&A...433..807M}.  The lines show the
  predictions of the Gaussian ``single-population'' model (i.e. a
  model in which the DTD is a narrow Gaussian centered at 3-4 Gyr;
  dashed line), \citep{2000ApJ...528..108Y} DD Chandrasekhar mass
  model (dotted line), and the ``two-populations'' shown in Figure~1
  (solid line).  The predictions use a Salpeter IMF and mass ranges of
  3-8M$_\odot$ \ (SN Ia) and 8-40M$_\odot$ \ (CC SNe), and are scaled
  to match the observed values.  Plot from
  \citet{2006MNRAS.370..773M}.  }
\label{ccratio}
\vskip 0.05 in
\end{figure}

{\it Future outlook:} There is a real possibility of determining the
progenitors of SNe Ia in the next decade, after combining accurate
$z<1$ rates with measures at $1<z<4$.  Different scenarios predict
different Ia to core collapse ratios with redshift
(Fig,~\ref{ccratio}).

Host galaxy photometry and spectroscopy can reveal the ages and
metallicity of the gas and stars \citep{2006ApJ...648..868S,
  2008ApJ...685..752G, 2008A&A...492..631A, 2009ApJ...691..661H}, and
DTDs can be directly constructed from this information
\citep{2008PASJ...60.1327T}.  Supernovae discovered at low redshift
can provide the greatest amount of host information, though
intermediate redshift surveys have a multiplex advantage when
multislit spectroscopy is used to build large samples.  These studies can benefit
by choosing well studied fields (e.g. COSMOS, GOODS, VVDS).

\section{Surviving or preceding material\label{stuff}}

Occasionally SNe Ia leave hints about the explosion process or
progenitors.  This can take the form of SN ejecta interacting with
previous phases of stellar mass loss \citep{2003Natur.424..651H}, or
absorption line evidence of mass loss episodes
\citep{2007Sci...317..924P}.  It is possible for progenitors to be
visible in pre-explosion X-ray images \citep{1997ARA&A..35...69K}.  
Finally, radio \citep{2006ApJ...646..369P} and optical
\citep{2007ApJ...670.1275L} searches for companion star material reach
contradictory conclusions unless the companion is also degenerate.

An exciting recent finding is the discovery of light
echoes enabling spectroscopic observations of historical SNe
\citep{2008ApJ...681L..81R, 2008Natur.456..617K}.  When combined with
knowledge of the SN remnant, this may allow us to connect SN
spectroscopic features with progenitor metallicity
\citep{2008arXiv0809.3151B, 2008ApJ...680L..33B}

{\it Future outlook:} While finding traces of progenitors is rare,
chances are directly proportional to the nearby SN discovery rate and
the aperture of the telescopes used.  X-ray facilities such
as Chandra, and in the future, IXO, will help to find possible
progenitors and study SN remnants in detail, and JWST and GSMTs may
directly pre-image progenitors in the optical or IR.  Rare signatures
of pre-SN mass loss require time series spectroscopic observations of
many supernovae to find the occasional goldmine.

\section{Galactic feedback and enrichment}
SNe Ia are a significant source of iron-peak elements and energy input
into the intergalactic medium.  The realization that a significant
fraction of SNe Ia occur only a few hundred million years after star
formation solved problems regarding cluster iron
abundances \citep[e.g.][]{2006MNRAS.372..265M,2005ApJ...629L..85S}.
X-ray observations of clusters can also constrain SN Ia rates and
models \citep{2007A&A...465..345D}.  

{\it Future outlook:} A refined understanding of SN Ia DTDs and theory
of energetics and elemental yields will allow progress
understanding galactic chemical enrichment.  The synergy between SN
rate studies and cluster studies, and between SNe Ia and X-ray studies
is again apparent.

\section{Theory and Explosion\label{theory}}
Theoretical explosion studies in 3d have only just begun
\citep{2005ApJ...623..337G}, though improvements in algorithms and
computing power should make such studies routine.  
New observations can constrain open theoretical questions about the explosion:
\begin{itemize}
\item IR spectra constrain the transition to detonation: lines from CI, OI, and MgII probe explosion products in
the outer layers \citep{2006ApJ...645.1392M}.  
\item Unburned carbon probes incomplete burning: \citep{2006Natur.443..308H, 2006ApJ...645.1392M, 2007ApJ...654L..53T,
  2007ApJ...669L..17H}.
\item Spectropolarimetry shows that some SNe Ia depart from spherical
symmetry \citep[see][]{2008ARA&A..46..433W}, forcing theoretical
creativity \citep{2004ApJ...616.1102W, 2007ApJ...660.1344R,
  2007A&A...465L..17H}.  
\item Late time observations place limits on $^{56}$Ni
production, the source of SN Ia luminosity \citep{2007Sci...315..825M}.
\item Oddball SNe Ia can require new classes of models \citep{2003Natur.424..651H, 2003PASP..115..453L}, even pushing classification boundaries
\citep{2006ApJ...653L.129B,2009arXiv0901.2074V}.  A few high
luminosity discoveries suggest the existence of super-Chandra mass
explosions \citep{2006Natur.443..308H}.
\end{itemize}

{\it Future outlook:} Theoretical studies most readily benefit from the
discovery of large numbers of low redshift SNe Ia, because this allows
more high S/N observations, the discovery of outliers,
spectropolarimetry, and challenging UV, IR, and late time
observations.

\section{Cosmology:  systematic errors}
\label{sys}
Arguably, the most important science done with SNe Ia is their use as
standard candles.  Systematic errors now dominate, but most are known
problems solvable with next-generation data sets.
Table 1 shows the dominant systematic errors for the
measurement of $w$, and we discuss the important ones here.

\begin{deluxetable}{llll}
\tablewidth{3.1in} \tablecaption{Current estimates of systematic errors on w}
\tablehead{
\colhead{Systematic} & \colhead{SNLS} & \colhead{ESSENCE} &\colhead{SDSS}
}
\startdata
Flux reference&0.053&0.02&0.037\\
Experiment zero points&0.01&0.04&0.014\\
Low-z photometry&0.02&0.005&\nodata\\
Landolt bandpasses&0.01&\nodata&0.019\\
Local flows&0.014&\nodata&0.04\\
Experiment bandpasses&0.01&\nodata&0.014\\
Malmquist bias model&0.01&0.02&0.017\\
Dust/Color-luminosity ($\beta$)&0.02&0.08&0.017\\
SN Ia Evolution&\nodata&0.02&\nodata\\
Restframe U band&\nodata&\nodata&0.08\\
\enddata
\tablecomments{ Systematic error estimates on $\langle w \rangle$ from
  \citet{2009conley}, \citet{2007ApJ...666..694W}, and
  \citet{2009kessler}.  \citet{2009arXiv0901.4804H} CfA3 systematics
  are similar to those for Wood-Vasey et al., though they are not
  separately tabulated.  The SDSS errors are for their MLCS2k2 fit.  
  Errors for each survey use their largest sample.  For the SNLS
  3rd year results the total systematic error is $\sim 0.06$,
  comparable to the statistical error, and the total statistical +
  systematic error is $\sim 0.09$.  The other studies find that
  systematic errors are dominant.
\label{tabsys}}
\end{deluxetable}

\subsection{Flux reference and zero points\label{calib}}
The dominant systematic in many studies (up to a 6\% systematic error
on $w$; Table~\ref{tabsys}), is flux calibration of photometry
across multiple passbands, an effect which introduces correlated
errors when comparing SNe at different redshifts.  Because of the need
to use historical low-z data, today SNe Ia are usually calibrated onto
the Landolt system, which is not sufficiently well understood and is
no longer fully reproducible.
A related problem is that at low redshift there is no single, uniform,
well characterized low $z$ data set, so we must cobble together inhomogeneous
data with varying qualities of flux calibration and bandpass
knowledge.  

{\it Future outlook:} The release of new low redshift data will allow
cosmological studies to move off of the Landolt system, though
selection effects, and the problems associated with combining data
sets will persist.  Further progress will require better calibration
in physical units, including the laboratory or in-situ measurements of
bandpasses.  The ACCESS experiment aims to establish a standard star
network based on physical units by calibrating a number of nearby
stars, e.g. Sirius, Vega, BD+174708 through direct comparison to NIST
standards.

A comparison low-redshift sample must be built for JDEM, including 500
SNe Ia at $z<0.1$ of sufficient flux calibration that they can be
compared to space-based detections \citep{2009arXiv0901.0721A}.  This
requires low redshift projects on the scale of this decade's
intermediate redshift efforts.

\subsection{The Ultraviolet}
The restframe ultraviolet accounts for the largest systematic in the
SDSS SN survey (Table 1). \citet{2006AJ....131..527J} indicate
that the restframe $U$-band is not as well behaved photometrically as
other optical bands, with as much as 0.08 mag dispersion.  Meanwhile,
it remains poorly understood spectroscopically,
\citep{2008ApJ...674...51E, 2008ApJ...684...68F}, possibly due to line
blanketing effects sensitive to metallicity.  Differences in
treatment of the restframe $U$-band account for many of the
differences in contemporary lightcurve fitters and directly impact their
extinction calculations.

{\it Future outlook:} Because of the difficulty of scheduling and
signal-to-noise requirements, space-based UV programs with HST and
Swift require the discovery of hundreds of nearby supernovae at very
early times over the course of a year.  This is only now becoming
possible with large low redshift surveys.  Another possibility
is studying SNe at $z\sim 0.2$ where the restframe $UV$ is shifted
into observed $g$-band.  SDSS-II will soon provide hundreds, and
ultimately LSST will provide thousands.  It may also be desirable to
avoid restframe $U$ in future studies like JDEM.

\subsection{Dust extinction}
\label{color}
Corrections for reddening due to dust in SNe Ia are complicated by the
fact that this correction is degenerate with an intrinsic
color-luminosity relation --- brighter SNe Ia are intrinsically
bluer, dimmer ones are redder
\citep{1996ApJ...473...88R,1999ApJ...525..209T}.  A further
complication is that the dust along the line of sight to SNe Ia does
not have the same average properties as Milky Way dust -- $R_B$
appears to range from 2-3, as compared to 4.1 for the average line of
sight in the Milky Way.  This is most apparent in low redshift studies
of SNe Ia with optical to IR photometry \citep{2007AJ....133...58K,
  2008ApJ...675..626W, 2008MNRAS.384..107E}, though it is also seen at
high redshift \citep{2006A&A...447...31A, 2007ApJ...664L..13C,
  2009arXiv0901.4804H}. 

Lightcurve fitters treat reddening differently.  SALT2
\citep{2007A&A...466...11G}, and SiFTO \citep{2008ApJ...681..482C} fit
for a color-luminosity relation with a slope, $\beta$, but do not
distinguish between the intrinsic SN dim-red relation and
dust.  MLCS2k2 \citep{2007ApJ...659..122J} attempts to separate the
two effects using assumptions and redshift-dependent priors.  Because
average corrections are made, either method is susceptible to the
observed evolution in supernova properties and environments with
redshift \citep{2007ApJ...667L..37H}.  

Estimates of the systematic impact on $w$ range from 0.02 to 0.08
(Table~\ref{tabsys}), arguably the dominant systematic.  At the
core of this error is a trifecta of ignorance: our lack of
understanding of dust in distant galaxies, our poor knowledge of the
intrinsic colors of SNe Ia, and our uncertainty regarding the
progenitor systems of SNe Ia and how the mix in SN Ia subtypes will
evolve with redshift.

{\it Future outlook: } It may be overly conservative to consider the
dust issue a 2\% in distance ``systematic floor'' as characterized by
the JDEM FoMSWG \citep{2009arXiv0901.0721A}, because there is hope for
solving it and possibly ways of circumventing it.

The reduced sensitivity to dust in the IR should allow the
characterization and mitigation of the dust extinction problem.  Long
wavelength baseline observations from the optical to the UV have
allowed determinations of the extinction law along the line of sight
to individual SNe \citep{2008MNRAS.384..107E, 2007AJ....133...58K}.
When huge SN samples are available in the next decade, indicators of
intrinsic SN color (e.g. certain spectroscopic features) may allow the
separation of intrinsic and dust reddening.  Meanwhile, SNe Ia appear
to be better standard candles in the near-IR, requiring little to no
lightcurve shape or color correction
\citep{2004ApJ...602L..81K,2008ApJ...689..377W}.  The sample size of
NIR spectra (now $>50$) is just beginning to make accurate
K-corrections feasible for the IR which will improve the calibration
of IR photometry.

While there is much room for growth in IR observations in the next
decade, they are expensive, particularly at high redshift.  Another
alternative is to construct dust-reduced samples of SNe Ia from those
in elliptical hosts, or far from the centers of galaxies.  This should
become possible with the discovery of hundreds to thousands more SNe
Ia in next-generation surveys.

\subsection{Evolution\label{evolution}}
There is strong evidence that some SNe Ia come from a short-lived
population of at most a few hundred million years (i.e. ``prompt''),
while ``tardy'' SNe Ia arise in an old population of at least several
Gyr \citep{2005A&A...433..807M, 2005ApJ...629L..85S,
  2006ApJ...648..868S}.  Moreover, the prompt SNe Ia are have broader
lightcurves and are on average brighter than their tardy counterparts
\citep{1996AJ....112.2391H,2001ApJ...554L.193H,2006ApJ...648..868S}.
Thus, as star formation increases by a factor of 10 from $z=0$ to
$z=1.5$, the ratio of prompt to tardy SNe Ia rises,
resulting in an increase in the average lightcurve width and intrinsic
brightness of SNe Ia \citep{2007ApJ...667L..37H}.  Additionally, SNe Ia at higher redshift
show weaker intermediate mass element features in
their spectra, consistent with the idea that they instead have more
iron-peak elements, including the $^{56}$Ni that gives rise to their
brighter lightcurves \citep{2009arXiv0901.2476S}.
\citet{2008ApJ...685..752G} and \citet{2009ApJ...691..661H} find that
metallicity has some effect on SN Ia $^{56}$Ni yield, and
thus luminosity.  

An evolving mix of SNe Ia need not inhibit cosmological
studies, because supernovae are calibrated based on their lightcurve
shape.  Problems could arise if the correction is imperfect
\citep{2009arXiv0901.4804H}, though \citet{2009ApJ...691..661H} find
that the SiFTO lightcurve fitter, at least, produces no Hubble
residuals with respect to the galaxy properties studied.

{\it Future outlook:} In the next decade it will be possible to
separate SN Ia samples by features like host galaxy metallicity, or
age of the stellar population.  Hints of demographic shifts will also
become well measured.  While many future SN Ia programs are looking
towards photometric identification of targets, rigorous testing of
possible SN Ia evolution with redshift requires spectroscopy.  The
host galaxies of SNe Ia should also be well characterized with
spectroscopy and UV to IR photometry to understand progenitor
populations.

\section{Priorities}
Many exciting science questions will be answered in the next decade by
the study of SNe Ia, including the nature of Dark Energy, but it will require
building large samples of supernovae from $0<z<4$, involving new resources.

The top priority for the immediate future is building new, well
controlled low redshift SN Ia samples comparable to those that
JDEM/EUCLID will ultimately produce.  Systematic effects are most
easily reduced by low redshift studies, and this is where most
non-cosmology SN Ia science is done.  New low redshift surveys to
discover hundreds of SNe per year starting in 2009/10 include PTF,
Skymapper, the La Silla SN search, and Pan-STARRS $3\pi$.  Skymapper
will produce its own lightcurves, but the rest will be limited by
follow-up resources.  Therefore, massive follow-up
programs targeting low-redshift SNe Ia are necessary.  One such
program is Las Cumbres Observatory Global Telescope Network (LCOGTN),
a planned worldwide network of 12-15 1 meter optical imaging
telescopes to supplement the existing 2 meter Faulkes North and South
telescopes.  A large fraction of the time on LCOGTN will be dedicated
to low redshift SN follow-up.  In addition, the
continued operation or growth of small, especially robotic or queue
scheduled telescopes is essential to the immediate future of SN Ia
science.  The NOAO ReSTAR initiative to grow the number of 2-4m
telescopes is of particular importance.  

Looking farther ahead, the next priority is measuring the Dark Energy
equation of state with JDEM/EUCLID, though only if there is a
significant SN component and a spectrograph that can resolve SN Ia
features (thousands of km/s).  In this case many of the selection
effects are mitigated by the single space mission concept, and the
follow-up resources are self-contained.

Of equal or greater importance to SN Ia studies is LSST.  This will be
the holy grail of transient studies, with tens of thousands of well
sampled multiband lightcurves per year.  However, many studies will be
limited by the follow-up resources available, e.g. spectroscopy, IR,
UV, polarimetry.

Finally, JWST, E-ELT, TMT, and GMT will allow great strides in SN Ia
science including the determination of SN Ia progenitors from high
redshift SN Ia rates, studies of evolution pushing to an era when SNe
Ia may be physically different, and allowing IR studies to become what
optical studies are today.

\bibliographystyle{apj}

\end{document}